\documentclass[10pt,journal]{IEEEtran}

\usepackage[pdftex]{graphicx}
\usepackage{algorithm}
\usepackage{color}
\usepackage{caption}
\usepackage{algpseudocode}

\definecolor{ForestGreen}{RGB}{162,52,0}

\usepackage{xcolor}
\usepackage{amsmath}
\usepackage{amssymb}
\usepackage{latexsym}
\usepackage{CJK}
\usepackage{url}
\usepackage{array}
\usepackage{ragged2e}
\usepackage{makecell}
\usepackage{multirow}
\usepackage{threeparttable}

\ifCLASSOPTIONcompsoc
  \usepackage[nocompress]{cite}
\else
  \usepackage{cite}
\fi

\ifCLASSINFOpdf
\else
\fi

\ifCLASSOPTIONcompsoc
  \usepackage[caption=false,font=footnotesize,labelfont=sf,textfont=sf]{subfig}
\else
  \usepackage[caption=false,font=footnotesize]{subfig}
\fi

\begin{document}

\title{Dynamic Virtual Network Embedding Algorithm based on Graph Convolution Neural Network and Reinforcement Learning}

\author{Peiying Zhang, Chao Wang, Neeraj Kumar,~\IEEEmembership{Senior Member,~IEEE}, \\ Weishan Zhang, and Lei Liu

\thanks{This work is supported in part by the National Natural Science Foundation of China under Grant 62001357, partially supported by the Major Scientific and Technological Projects of CNPC under Grant ZD2019-183-006, partially supported by the Guangdong Basic and Applied Basic Research Foundation under Grant 2020A1515110496 and 2020A1515110079, and partially supported by Shandong Provincial Natural Science Foundation, China under Grant ZR2020MF006. \textit{(Corresponding authors: Neeraj Kumar and Peiying Zhang)}.}
\thanks{Peiying Zhang, Chao Wang and Weishan Zhang are with the School of Computer Science and Technology, China University of Petroleum (East China), Qingdao 266580, China. E-mails: zhangpeiying@upc.edu.cn, wangch\_upc@qq.com and zhangws@upc.edu.cn.}
\thanks{Neeraj Kumar is with the Department of Computer Science and Engineering, Thapar Institute of Engineering and Technology, Deemed to be University, Patiala 147004, India, also with the Department of Computer Science and Information Engineering, Asia University, Taichung 41354, Taiwan, and also with School of Computer Science, University of Petroleum and Energy Studies, Dehradun Uttarakhand, India. E-mail: neeraj.kumar@thapar.edu.}
\thanks{Lei Liu is with State Key Laboratory of Integrated Services Networks, Xidian University, Xi'an 710071, China. Email: leiliu@xidian.edu.cn.}
}

\markboth{IEEE Internet of Things Journal,~Vol.~XX, No.~XX, XX~2021}
{}

\maketitle
\begin{abstract}
Network virtualization (NV) is a technology with broad application prospects. Virtual network embedding (VNE) is the core orientation of VN, which aims to provide more flexible underlying physical resource allocation for user function requests. The classical VNE problem is usually solved by heuristic method, but this method often limits the flexibility of the algorithm and ignores the time limit. In addition, the partition autonomy of physical domain and the dynamic characteristics of virtual network request (VNR) also increase the difficulty of VNE. This paper proposed a new type of VNE algorithm, which applied reinforcement learning (RL) and graph neural network (GNN) theory to the algorithm, especially the combination of graph convolutional neural network (GCNN) and RL algorithm. Based on a self-defined fitness matrix and fitness value, we set up the objective function of the algorithm implementation, realized an efficient dynamic VNE algorithm, and effectively reduced the degree of resource fragmentation. Finally, we used comparison algorithms to evaluate the proposed method. Simulation experiments verified that the dynamic VNE algorithm based on RL and GCNN has good basic VNE characteristics. By changing the resource attributes of physical network and virtual network, it can be proved that the algorithm has good flexibility.
\end{abstract}


\IEEEpeerreviewmaketitle

\section{Introduction}\label{pa1}

The fact that the traditional Internet architecture is becoming rigid has promoted the development of future network technologies \cite{a1}. Network virtualization (NV) will realize the effective management of the next generation Internet and become an important part of the future network architecture \cite{z1,a2}. NV mainly realizes the decoupling of network services, functions and specific underlying hardware, which greatly improves the flexibility and scalability of the underlying network architecture. NV involves two main roles: infrastructure provider (InP) and service provider (SP) \cite{a3}. The process that the two cooperate to allocate physical network resources for virtual network requests (VNRs) is called virtual network embedding (VNE). The ideal result of VNE is to reasonably schedule the underlying network resources to receive as many VNRs as possible, improve the acceptance rate of virtual network, realize load balancing, maximize resource consumption revenue and reduce costs \cite{z2,a4}. Radio network resource management faces severe challenges, including storage, spectrum, computing resource allocation, and joint allocation of multiple resources \cite{jcx1,jcx2}. With the rapid development of communication networks, the integrated space-ground network has also become a key research object \cite{jcx3}.

VNE is a typical NP-hard problem, which determines that the VNE problem can only obtain an approximate optimal solution instead of an optimal solution \cite{a5,a6}. Usually, heuristic method is used to solve the VNE process, or VNE is modeled as a combinatorial optimization problem, and then (integer) linear programming is used to solve it. The above methods are traditional solutions to VNE problem, which play a positive role in a certain social application and technical background. The analysis shows that there are several common disadvantages in traditional VNE solutions.

(1) The constraints and optimization objectives of VNE are made manually. Unnatural conditional control will damage the robustness and flexibility of VNE algorithm.

(2) VNE algorithm with differentiated quality of service (QoS) may have high real-time requirement, so it is time sensitive. Existing algorithms usually ignore this problem.

(3) Ignore the dynamic characteristics of VNE, i.e., during the VNR cycle, the resource configuration of virtual nodes and links, the virtual topological structure, and the number of physical network resources are all in dynamic changes.

The application of machine learning (ML) algorithms in VNE has achieved great success \cite{z3,a7}. Through the efficient interaction between learning agent and environment, RL has excellent decision-making capabilities. At each time step $t$, the intelligent agent interacts with the environment to obtain the environment state $s_t$. After that, the agent will apply an action $a_t$ to the environment to transfer the state of the environment to $s_{t+1}$, and at the same time, the environment will feed back to the agent a reward $r_t$. The goal of RL is to help the agent maximize the reward signal. The detailed application of RL will be introduced below. The VNE algorithm based on RL has been well experimented, so RL can be applied to our work.

Problems such as node ordering and pattern matching can all be simplified by graph-related theories and techniques. Graph neural network (GNN) is a new type of ML model architecture that can aggregate graph features (degrees, distance to specific nodes, node connectivity, etc.) on nodes \cite{a8}. The model can be used to cluster nodes and links according to the physical nodes and physical links attribute characteristics (CPU, storage, bandwidth, delay, etc.) and search for the best VNE strategy by intelligent agent training. Graph convolution neural network (GCNN) is a natural generalization form of GNN, which is highly suitable for graph structures of any topological form \cite{a9}. Therefore, the organic combination of VNE and GCNN has a good prerequisite. To this end, The main work of this paper is as follows:

(1) A VNE algorithm based on RL and GCNN is proposed. In particular, GCNN is used to automatically extract the features of underlying network, which optimizes the selection of VNE decision.

(2) We notice the dynamic characteristics of VNE process, and adopt a dynamic VNE method based on fitness value to deal with the changing virtual network topology and resource attributes.

(3) We conducted a large number of simulation experiments and compared the typical indicators of VNE with other algorithms to verify the effectiveness of the algorithm. In addition, the flexibility of the algorithm is verified by changing the attribute parameter distribution of physical network and VNRs.

The rest of this paper is arranged as follows. Section \ref{pa2} introduces the related work of VNE. Section \ref{pa3} gives the necessary network model to solve VNE problem and describes the related technical problems. Section \ref{pa4} formulates the related problems of VNE. Section \ref{pa5} introduces the design and implementation details of the algorithm. Section \ref{pa6} verifies and analyzes the performance of the algorithm through simulation experiments. In the last section, we summarize the whole paper.

\section{Related Work}\label{pa2}

Heuristic method is the traditional method for solving VNE process, including meta heuristic algorithm, mathematical optimization and so on. VNE solutions based on ML often use algorithms or models such as deep learning (DL), RL or graph theory, which has gradually become the mainstream method to solve NP-hard problems such as VNE \cite{z4}.

\subsection{VNE Solution based on Heuristic}

The more classic is the node-link two-stage coordinated and embedded ViNEYard algorithm proposed by Chowdhury et al. \cite{1}. This algorithm modeled VNE as a mixed-integer linear programming problem, and then implemented D-ViNE and R-ViNE using deterministic and random methods respectively. The results showed that these algorithms can effectively improve the revenue of VNE. Increasing the VNE revenue or reducing the embedding cost is an important goal of VNE algorithm. Based on Markov random walk model theory, Cheng et al. \cite{2} sorted the importance of network nodes according to network topology attributes, and proposed two representative VNE algorithms based on node sorting. According to whether the virtual link was divided into paths, it was decided to use the multi-commodity flow algorithm or the breadth-first search algorithm to perform the virtual link mapping process. These two algorithms had effectively improved the long-term average revenue, and the method based on node sorting has also become a typical and popular method in VNE. Zhang et al. \cite{3} first considered VNE algorithm with multi-dimensional resource constraints of computing, network and storage, and proposed two heuristic algorithms NRM-VNE and RCR-VNE as baseline algorithms.

Since VNE can be modeled as a search problem in a larger three-dimensional space, meta heuristic method can be used as an important means to solve VNE. Meta heuristic algorithm is an improvement of heuristic algorithm, which adds local search on the basis of heuristic algorithm \cite{m2}. Particle swarm algorithm (PSO) is a meta heuristic algorithm commonly used to solve VNE \cite{4,5,m1}. Different VNE strategies can be used as particles to perform a global optimal search based on the objective function set in advance.

As we mentioned earlier, the heuristic method of manually formulating constraints and features for VNE algorithm greatly limits the flexibility of VNE algorithm. Moreover, with the rapid expansion of physical network and user scale, the global search strategy efficiency of heuristic algorithm gradually decreases, and there is a big gap between the time performance and the latest VNE algorithm.

\subsection{VNE Solution based on ML}

With the successful practice of ML algorithms in real life, scholars began to explore and tried to apply ML algorithms in VNE, among which RL algorithm was the most representative \cite{6}. Haeri et al. \cite{7} modeled VNE as a Markov decision process, and then used Monte Carlo search tree (MCST) algorithm to implement the node mapping process. In the stage of virtual link embedding, the authors used multi-commodity flow algorithm to implement the MaVEn-M algorithm, and the shortest path algorithm to implement the MaVEn-S algorithm. The profitability and profitability of InP illustrated the effectiveness of the algorithm. It cannot be ignored that MCST algorithm should perform a complete VNE process every time the virtual nodes are embedded, which makes the algorithm's time cost higher. Q-learning is another commonly used RL model. Yuan et al. \cite{8} implemented the VNE algorithm based on this model to find the optimal VNE strategy for RL agent, in which the Q-matrix and the Q-table played an important role. One problem with this algorithm is that the extraction of virtual network features is done manually. The efficiency of the algorithm will decrease when the VNR scale is large or the structure is complex. Yao et al. \cite{9} implemented a VNE algorithm based on single layer policy network. According to the historical embedded data of virtual network, RL agent used strategy gradient descent to find the optimal embedding strategy.

The authors of \cite{10} and \cite{11} also combined DL and RL, and their deep reinforcement learning (DRL) based VNE algorithm also provided new ideas for solving this problem.

In the field of ML, VNE algorithm based on graph theory has gradually come into people's view. Habibi et al. \cite{12} creatively combined GNN and VNE, and then proposed GraphViNE algorithm based on graph automatic encoder. The biggest feature of this algorithm is that servers with similar resources are clustered, which ultimately effectively reduces the running time of the algorithm. Yan et al. \cite{13} paid attention to the time constraint of VNE and applied GNN, especially GCNN, to VNE. It was worth noting that the authors proposed a parallel framework and multi-objective reward function based on DRL, which took into account VNE revenue, acceptance rate and load balance. Reference \cite{14} and \cite{15} respectively proposed VNE algorithm based on graph feature space alignment and VNE algorithm based on subgraph matching. The former mainly associated physical nodes with virtual nodes, while the latter utilized a hierarchical physical management architecture. Both of them are good applications of graph theory in VNE problem.

The above research does not support the embedding of dynamic virtual networks well, which will greatly affect the utilization of resources. Dehury et al. \cite{16} focused on the dynamic characteristics of virtual network topology, and proposed a dynamic VNE algorithm based on fitness value. The function of fitness value was to mark the utilization degree of physical resources. At the same time, the authors used the method of studying part of physical servers to replace the method of studying all physical servers, which effectively improved the efficiency of the algorithm. In addition, references \cite{17} and \cite{18} are also excellent representatives of dynamic VNE algorithm. But the analysis finds that they do not fully apply ML methods.

\section{Network Model and Problem Description}\label{pa3}

For easy viewing, we summarize the commonly used network representation symbols in TABLE \ref{tab_1}.

\begin{table}
\centering
\caption{Notation of Network Model}
\renewcommand\arraystretch{1.5}
\begin{tabular}{|p{10mm}|p{40mm}|}
\hline
Notation & Description  \\
\hline
$G^P$ & physical network \\
$N^P$ & physical node \\
$L^P$ & physical link \\
$A^N$ & physical node resource attributes \\
$A^L$ & physical link resource attributes  \\
\hline
$G_i^V$ & the $i$-th virtual network request \\
$N^V$ & virtual node \\
$L^V$ & virtual link \\
$R^V$ & virtual node resource requirements \\
$R^L$ & virtual link resource requirements \\
\hline
\end{tabular}
\label{tab_1}
\end{table}

\subsection{Physical Network and VNR Modeling}

Physical network can be regarded as a weighted undirected graph $G^P=\{N^P,L^P,A^N,A^L\}$. $N^P$ represents the node set composed of all underlying physical nodes, and $L^P$ represents the link set composed of all physical links. $A^N$ and $A^L$ represent physical node attribute set and physical link attribute set respectively. We mainly regard CPU resources as physical node attributes and bandwidth resources as physical link attributes, namely $A^N=\{c_{n_1^p},c_{n_2^p},...,c_{n_{|N^P|}^p}\}$ and $A^L=\{b_{l_1^p},b_{l_2^p},...,b_{l_{|L^P|}^p}\}$, where $|N^P|$ and $|L^P|$ represent the total number of physical nodes and physical links, respectively.

Similarly, virtual network is regarded as a weighted undirected graph $G_i^V=\{N^V,L^V,R^N,R^L\}$, and $G_i^V$ indicates that this is the $i-th$ virtual network. $N^V$ represents the set of all virtual nodes, and $L^V$ represents the set of all virtual links. $R^N$ and $R^L$ respectively represent the resource requirement attribute set of virtual node and virtual link, where $N^V=\{c_{n_1^v},c_{n_2^v},...,c_{n_{|N^V|}^v}\}$ represents the CPU resource requirement of virtual node, and $L^V=\{b_{b_1^v},b_{b_2^v},...,b_{b_{|L^V|}^v}\}$ represents the bandwidth resource requirement of virtual link. $|N^V|$ and $|L^V|$ respectively represent the number of virtual nodes and virtual links in the virtual network.

\subsection{VNE Model and Description}

\begin{figure*}[!h]
\centering
\includegraphics[width=0.85\textwidth]{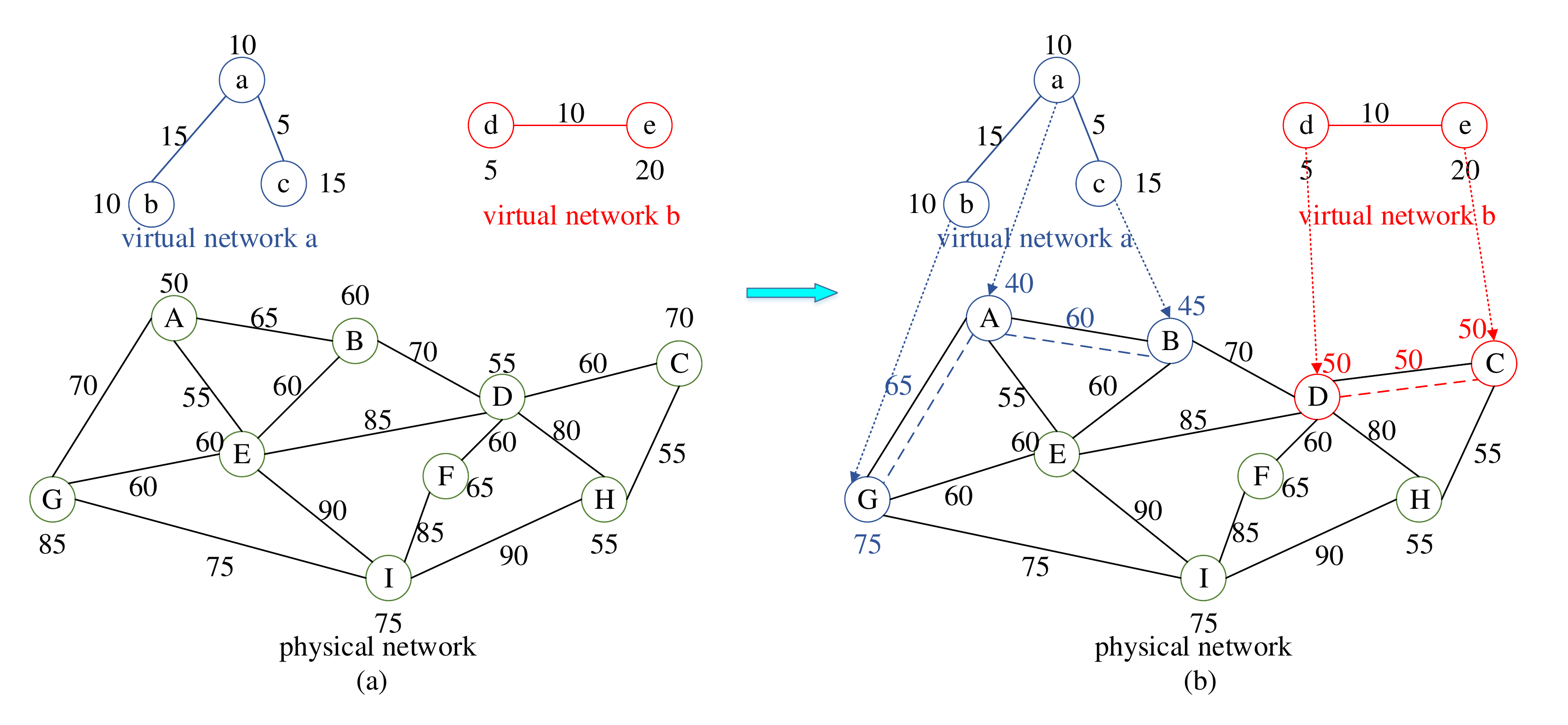}
\caption{Schematic diagram of physical network and virtual network. (a) VNRs before embedding. (b) A possible result of VNRs embedding.}
\label{fig_1}
\end{figure*}

Based on physical network and virtual network model, the VNR can be expressed as $VNR_i=(G_i^V,t_a,t_e)$, where $t_a$ represents the time when the VNR arrives, and $t_e$ represents the time when the VNR leaves. The VNE process can be modeled as $G_i^V(N^V,L^V) \uparrow G^{P'}(N^{P'},L^{P'})$, where $G^{P'}(N^{P'},L^{P'})$ represents a partial subgraph of physical network. Thus, a complete VNE process includes two stages: node mapping and link mapping. Fig. \ref{fig_1} gives a visual representation of the network topology, and shows the result of two VNRs embedded in the physical network. We mark the CPU resource demand and bandwidth resource demand respectively next to network node and link. The target node and link should have sufficient resources to carry them.

\subsection{Dynamic VNE}

Changes in the number of network users or fluctuations in workload will cause the structural change of virtual network. The most intuitive impact of changes in the number of users is the changes in virtual nodes and links, i.e., changes in structural configuration. Workload fluctuations usually change the number of resource requirements of virtual nodes or links, i.e., the resource configuration changes. The above changes are real in the process of VNE. The virtual network whose structure configuration or resource configuration changes with time interval is called dynamic virtual network, and the process of mapping dynamic virtual network to physical network is called dynamic VNE.

\begin{figure}[!h]
\centering
\includegraphics[width=1\columnwidth]{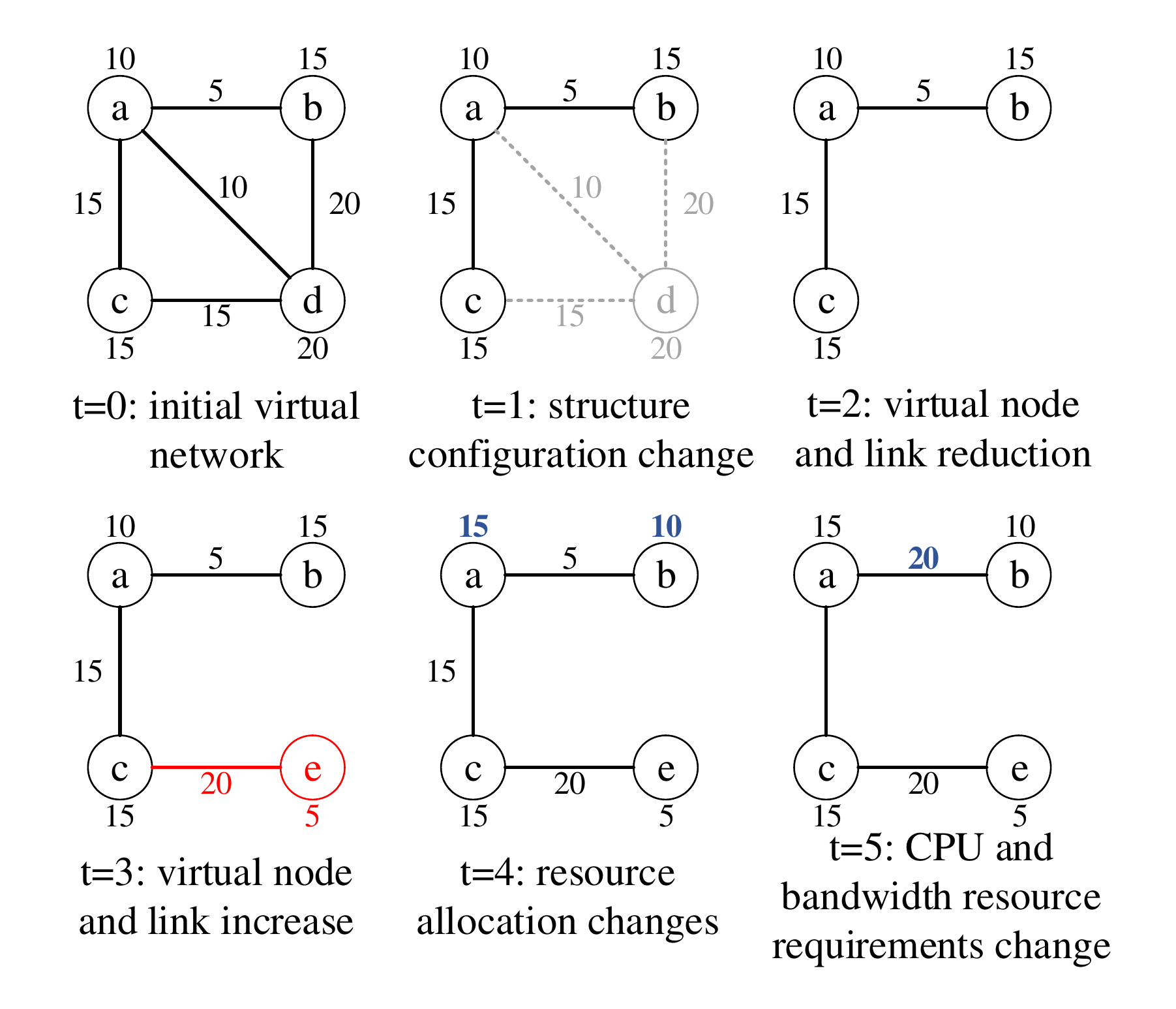}
\caption{Dynamic virtual network example.}
\label{fig_2}
\end{figure}

Fig. \ref{fig_2} shows the change process of virtual network structure configuration and resource configuration in different time intervals. $t=0$ to $t=5$ is a continuous VNE process. When $t=0$, the virtual network structure is the initial form, and the resource requirements of virtual nodes and links are shown in the figure. When $t=1$, the configuration of virtual network is changed, and virtual node $d$ and its related virtual links $l^v(a,d)$, $l^v(b,d)$, $l^v(c,d)$ are deleted. When $t=2$, the structural configuration change process of virtual network ends, and a new virtual network topology with three virtual nodes and two virtual links is formed. When $t=3$, the configuration of virtual network changes again, and a new virtual node $e$ and a virtual link $l^v(c,e)$ are added. When $t=4$, the resource configuration of virtual network has changed. The number of CPU resource requirements for node $a$ increases from 10 units to 15 units, and the number of CPU resource requirements for node $b$ decreases from 15 units to 10 units. When $t=5$, the number of bandwidth resources of virtual link $l^v(a,b)$ also changes, from the original 5 units to 20 units.

\section{Problem Formulation}\label{pa4}

\subsection{Restrictions}

We use the binary variable $\varepsilon_{n_k^v}^{n^p}$ to measure whether the physical node $n^p$ is embedded by the virtual node $n_k^v$, where $k$ is the virtual node label, $k=1,2,...,|N_i^V|$. $|N_i^V|$ represents the number of virtual nodes in virtual network $G_i^V$. Then $\varepsilon_{n_k^v}^{n^p}$ can be expressed as follows,
\begin{equation}
\begin{aligned}
\varepsilon_{n_k^v}^{n^p} = &
\left\{
             \begin{array}{ll}
             1, & if\,\,n_k^v \uparrow n^p, \\
             0, & others.
             \end{array}
\right.
\end{aligned}
\end{equation}

Another binary variable $\delta_{l_k^v}^{l^p}$ is used to measure whether the virtual link $l_k^v$ is mapped to the physical link $l^p$, where $k$ is the virtual link label, $k=1,2,...,|L_i^V|$. $|L_i^V|$ indicates the number of virtual links in virtual network $G_i^V$. Then $\delta_{l_k^v}^{l^p}$ can be expressed as follows,
\begin{equation}
\begin{aligned}
\delta_{l_k^v}^{l^p} = &
\left\{
             \begin{array}{ll}
             1, & if\,\,l_k^v \uparrow l^p, \\
             0, & others.
             \end{array}
\right.
\end{aligned}
\end{equation}

VNE needs to follow resource, location and access control constraints. Here we mainly set the resource constraints of VNE. The specific constraint content is shown in the following formula.

\begin{equation}
\begin{aligned}
\sum\limits_{k=1}^{|N_i^V|}\varepsilon_{n_k^v}^{n^p}=1,n_k^v \in G_i^V,
\end{aligned}
\end{equation}
\begin{equation}
\begin{aligned}
\sum\limits_{k=1}^{|L_i^V|}\delta_{l_k^v}^{l^p}\geq1,l_k^v \in G_i^V,
\end{aligned}
\end{equation}
\begin{equation}
\begin{aligned}
A\_{c_n^p}=c_{n^p}-\sum\limits_{k=1}\varepsilon_{n_k^v}^{n^p} \cdot c_{n_k^v},c_{n_k^v}\,\,from\,\,different\,\,G^V,
\end{aligned}
\end{equation}
\begin{equation}
\begin{aligned}
A\_{b_l^p}=b_{l^p}-\sum\limits_{k=1}\delta_{l_k^v}^{l^p} \cdot b_{l_k^v},b_{l_k^v} \in G_i^V,i=1,2,...,|VNR|,
\end{aligned}
\end{equation}
\begin{equation}
\begin{aligned}
c_{n_k^v} \leq A\_{c_n^p},if\,\,\varepsilon_{n_k^v}^{n^p}=1,
\end{aligned}
\end{equation}
\begin{equation}
\begin{aligned}
b_{l_k^v} \leq A\_{b_l^p},if\,\,\delta_{l_k^v}^{l^p}=1.
\end{aligned}
\end{equation}

Formula (3) shows that virtual nodes $n_k^v$ from the same virtual network $G_i^V$ can only be mapped to different physical nodes. Formula (4) shows that a virtual link $l_k^v$ may be mapped to multiple physical links. Formulas (5) and (6) represent the current available resource capacities of physical nodes $n^p$ and links $l^p$, respectively. Formula (7) and formula (8) indicate that when virtual node $n_k^v$ and virtual link $l_k^v$ are embedded on physical node $n^p$ and physical link $l^p$, respectively, the resource capacity of $n^p$ and $l^p$ is sufficient.

\subsection{Evaluation Index}

The classical evaluation indexes of VNE algorithm include long-term average revenue, resource consumption cost, resource utilization rate, virtual network acceptance rate and revenue-cost ratio. Among them, the revenue of VNE are measured by the consumption of CPU resources and bandwidth resources,
\begin{equation}
\begin{aligned}
R(G_i^V)=\sum\limits_{n^v \in N^V}c_{n^v}+\sum\limits_{l^v \in L^V}b_{l^v},N^V,L^V \in G_i^V,
\end{aligned}
\end{equation}
where $c_{n^v}$ represents the CPU resource requirement of virtual node $n^v$ and $b_{l^v}$ represents the bandwidth resource requirement of virtual link $l^v$.

The resource consumption cost of VNE is expressed as follows,
\begin{equation}
\begin{aligned}
C(G_i^V)=\sum\limits_{n^v \in N^V}c_{n^v}+\sum\limits_{l^v \in L^V}\sum\limits_{l^p \in L^P}b_{l^v}^{l^p}, \\ N^V,L^V \in G_i^V, L^P \in G^P.
\end{aligned}
\end{equation}

The main difference between the two is reflected in the consumption of bandwidth resources. Since a virtual link may be mapped to multiple physical links due to path segmentation, the cost of bandwidth consumption will increase, and the revenue of bandwidth consumption only depends on its own bandwidth resource demand. Based on both, the long-term average revenue of VNE is defined as follows,
\begin{equation}
\begin{aligned}
R=\lim_{T \to \infty}\frac{\sum\limits_{t=0}^T\sum\limits_{i=1}^{|VNR|}R(G_i^V,t)}{T},
\end{aligned}
\end{equation}
where $\sum\limits_{t=0}^T\sum\limits_{i=1}^{|VNR|}R(G_i^V,t)$ refers to the total revenue of VNR embedding in the time range $T$.

The VNE revenue-cost ratio is defined as follows,
\begin{equation}
\begin{aligned}
R/C=\lim_{T \to \infty}\frac{\sum\limits_{t=0}^T\sum\limits_{i=1}^{|VNR|}R(G_i^V,t)}{\sum\limits_{t=0}^T\sum\limits_{i=1}^{|VNR|}C(G_i^V,t)},
\end{aligned}
\end{equation}
where $\sum\limits_{t=0}^T\sum\limits_{i=1}^{|VNR|}C(G_i^V,t)$ refers to the resource cost consumed by the VNR embedding in the time range $T$.

The VNR acceptance rate is defined as follows,
\begin{equation}
\begin{aligned}
AR=\lim_{T \to \infty}\frac{\sum\limits_{t=0}^T |VNR|_{acc}}{\sum\limits_{t=0}^T |VNR|_{arr}},
\end{aligned}
\end{equation}
where $|VNR|_{acc}$ is the number of VNRs successfully mapped to the physical network in the VNE process. $|VNR|_{arr}$ is the total number of VNRs that come in the VNE process.

\subsection{Fitness Matrix and Value}

There needs to be a standard to record the degree of utilization of CPU resources, and the fitness matrix can play a role. The value in the matrix is the remaining CPU capacity, and the calculation method is shown in formula (5). We use $f_{ij}(t)=A\_c_{n^p}$ to mark every element in the fitness matrix, where $i$ represents the physical node in each row, and $j$ represents the virtual node in each column. Therefore, the dimension of fitness matrix depends on the number of virtual nodes embedded at time $t$. The sign $F(t)=f_{ij}(t)$ is used to refer to the fitness matrix. We specify that the elements of fitness matrix are recorded as follows,
\begin{equation}
\begin{aligned}
f_{ij}(t) = &
\left\{
             \begin{array}{ll}
             \infty, & f_{ij} \leq 0, \\
             A\_{c_n^p}, & others.
             \end{array}
\right.
\end{aligned}
\end{equation}

We define the concept of fitness value as the availability of physical nodes to characteristic virtual nodes. The calculation method is as follows,
\begin{equation}
\begin{aligned}
\alpha_j(t)=\sum\limits_{i=1}^{|N^P|}\varepsilon_{n{j^v}}^{n_i^p} \cdot \frac{f_{ij}(t)}{c_{n^p}}.
\end{aligned}
\end{equation}

We use an adjoint matrix of a fitness matrix to mark whether there is a link between any two physical nodes. The rows and columns of the adjoint matrix represent the physical node numbers, and the value of each element in the matrix is defined as follows,
\begin{equation}
\begin{aligned}
Ad_{ij}(t) = &
\left\{
             \begin{array}{ll}
             b_{l^p}, & l^p \,\, between \,\, n_i^p \,\, and \,\, n_j^p, \\
             \infty, & no \,\, link \,\, between \,\, n_i^p \,\, and \,\, n_j^p.
             \end{array}
\right.
\end{aligned}
\end{equation}

Considering CPU resources and bandwidth resources, the overall fitness value of the algorithm is calculated as follows,
\begin{equation}
\begin{aligned}
\beta(t)=\sum\limits_{j=1}^{num\_n^v}\alpha_j(t) + \sum\limits_{i=1}^{|L^P|}\sum\limits_{j=1}^{num\_l^v}\delta_{l_j^v}^{l_i^p} \cdot \frac{A\_b_{l^p}}{b_{l^p}},
\end{aligned}
\end{equation}
where $num\_n^v$ and $num\_l^v$ respectively represent the total number of virtual nodes and virtual links successfully mapped in the entire VNR process.

The ultimate goal of the algorithm is,
\begin{equation}
\begin{aligned}
& Minimize\sum\limits_{j=1}^{num\_n^v}\min\limits_{1 \leq i \leq |L^P|}(\varepsilon_{n_j^v}^{n_i^p} \cdot \frac{A\_c_{n^p}}{c_{n^p}}+ \delta_{l_j^v}^{l_i^p} \cdot \frac{A\_b_{l^p}}{b_{l^p}}) \\
& = Minimize\sum\limits_{j=1}^{num\_n^v}\min\limits_{1 \leq i \leq |L^P|}(\varepsilon_{n_j^v}^{n_i^p} \cdot \frac{f_{ij}(t)}{c_{n^p}}+ \delta_{l_j^v}^{l_i^p} \cdot \frac{A\_b_{l^p}}{b_{l^p}}) \\
& = Minimize \, \beta(t),
\end{aligned}
\end{equation}
where $\varepsilon_{n_j^v}^{n_i^p}$ is used to measure whether the virtual node $n_j^v$ is mapped to the physical node $n_i^p$, and $\delta_{l_j^v}^{l_i^p}$ is used to measure whether the virtual link $l_j^v$ is mapped to the physical link $l_i^p$.

\section{Algorithm Design and Implementation}\label{pa5}

RL mainly includes four entities: intelligent agent, state $s_t$, action $a_t$ and reward $r_t$. When RL is applied to VNE problem, we use a self defined GCNN model as RL agent, which can extract network features for training, and then get the VNE strategy. The specific content of GCNN model will be introduced in the following. $s_t$ refers to the information that agent can obtain from virtual network environment. We define $s_t$ to include the available CPU resources of each physical node, the available bandwidth resources of each physical link, the resource requirements of each virtual node and each virtual link, the virtual nodes and virtual links that have been successfully mapped to physical network in the current VNR, and the virtual nodes and virtual links that have not been mapped in the current VNR. $a_t$ refers to a VNE strategy adopted by agent, which includes node mapping strategy and link mapping strategy. $r_t$ is a reward function that agent gets by adopting some embedding strategy to the environment. It should be noted that RL agent has no strict objective function or label. It accumulates reward signals through continuous training. We define the calculation method of $r_t$ obtained by $a_t$ as follows,
\begin{equation}
\begin{aligned}
r(a_t)=R(G_i^V) \cdot R/C.
\end{aligned}
\end{equation}

Convolutional neural network (CNN) can automatically extract the advanced features of images. It has been shown that CNN can provide a great help in dealing with VNE related problems \cite{6,9}. But the original intention of CNN design is to solve the Euclidean graph with elements arranged in order, and the efficiency will be much lower when dealing with any graph structure such as network topology. Reference \cite{19} proposed a method of convolution operation on random graph topology under the background of spectrograph theory, i.e., GCNN method. The essence of GCNN is Fourier transform. The key to the problem is to find a Laplacian in the graph and use an orthogonal factor to characterize the network topology. The following Fourier transform form can be used,
\begin{equation}
\begin{aligned}
F(\lambda_v)=\hat{f}(\lambda_v)=\sum\limits_{i=1}^Nf(i)\mu_v^*(i),
\end{aligned}
\end{equation}
where $f$ represents the vector, $N$ represents the dimension of the vector, $f(i)$ corresponds to the network node, and $\mu_v^*(i)$ represents the $i$-th component of the $v$-th feature vector. The Fourier change of $f$ is the inner product operation of the feature vector $\mu_v^*(i)$ corresponding to the feature value. Namely,
\begin{equation}
\begin{aligned}
\hat{f}=U^Tf.
\end{aligned}
\end{equation}

Perform the inverse Fourier transform to get,
\begin{equation}
\begin{aligned}
f=U^T\hat{f}.
\end{aligned}
\end{equation}

According to convolution theorem, the Fourier transform of function convolution is the product of function Fourier transform, then the Fourier transform of $f$ and convolution kernel $g$ on the network topology diagram is defined as follows,
\begin{equation}
\begin{aligned}
f*g=\Gamma^{-1}(\hat{f}(\omega).\hat{g}(\omega))=\frac{1}{2\pi}\int\hat{f}(\omega).\hat{g}(\omega)e^{-iwt}d\omega,
\end{aligned}
\end{equation}
where $g=\sum_{k=0}^K\alpha_k\Lambda^k$, $\alpha_k$ is the training parameter and $\Lambda$ is the diagonal matrix with eigenvalue $\lambda_v$. Therefore, the final output of GCNN is as follows,
\begin{equation}
\begin{aligned}
g\_o=\sum\limits_{k=0}^K\alpha_k\lambda_vf.
\end{aligned}
\end{equation}

\begin{figure*}[!h]
\centering
\includegraphics[width=0.78\textwidth]{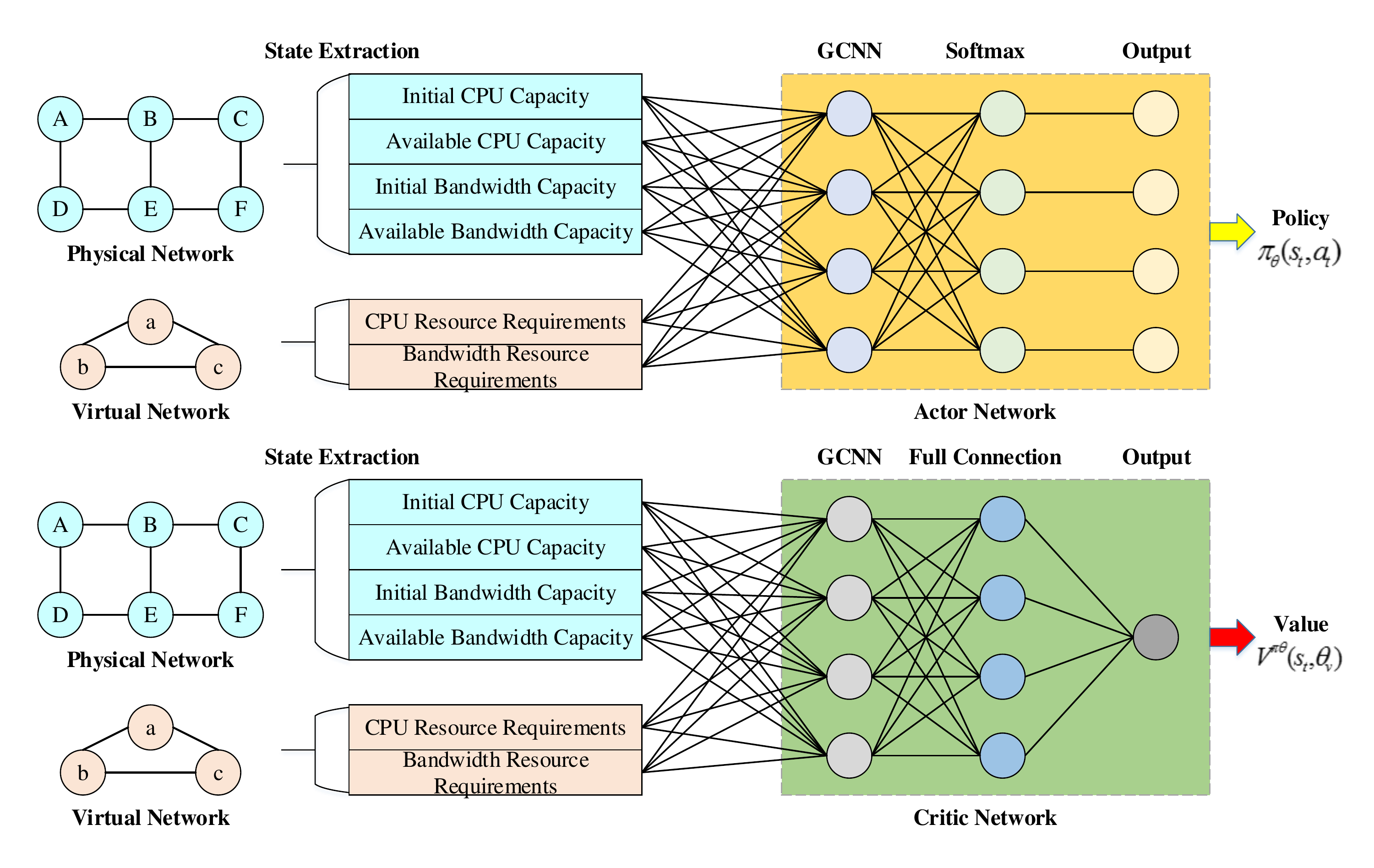}
\caption{Graph convolution neural network model based on Reinforcement Learning.}
\label{fig_3}
\end{figure*}

We organize the topological features extracted by GCNN into a column vector, and make it output a probability distribution equal to the number of physical network nodes through the softmax operation. Then we use the asynchronous advantage actor-critic algorithm to optimize the parameters of neural network. The algorithm includes an actor network and a critic network. The former is used to generate a set of parameterized strategies $\pi^\theta$, and the latter is used to generate a set of parameterized estimated values $v^{\pi\theta}(s_t,\theta_v)$. Considering the important processes of feature extraction, strategy generation and model training, the GCNN model based on RL is shown in Fig. \ref{fig_3}.

The execution of dynamic VNE requires a predetermined duration $D$ of the entire VNE. The process of dynamic VNE is shown in Algorithm 1.

\begin{algorithm}[h]
  \caption{Dynamic VNE algorithm}
  \begin{algorithmic}[1]
    \Require
    {$G^P(N^P,L^P),G^V(N^V,L^V),D$};
    \State
    {$Calculate\,\,fitness\,\, value\,\, by\,\, formulas\,\, (15)\,\, and\,\, (16)$};
    \State
    {$flag=FALSE$};
    \While {$flag=FALSE$}
    \State {$flag=TRUE$};
    \For {each $n_i^v \in N^V \& n_j^p \in N^P$}
    \State {$n_i^v \uparrow n_j^p$};
    \If {$another\,\, n_k^v \uparrow n_j^p$}
    \State {$n^v = n_i^v + n_k^v$};
    \State {$Calculate\,\,new\,\,fitness\,\, value\,\,of\,\,n^v$};
    \Else
    \State {$Calculate\,\,new\,\,fitness\,\, value\,\,of\,\,n_k^v$};
    \EndIf
    \If {$\beta(t) \geq \beta(t)'$}
    \State {$flag=TRUE$};
    \EndIf
    \EndFor
    \EndWhile
    \While {$D \ge 0$}
    \State {$D -= 1$};
    \State {$Perform\,\,a\,\,virtual\,\,network\,\,embedding\,\,process$};
    \EndWhile
  \end{algorithmic}
\end{algorithm}

The asynchronous advantage actor-critic algorithm is the key process of realizing RL agent training. This algorithm uses a model parallel training method, which can effectively improve the execution efficiency and convergence of the algorithm. The training process of the asynchronous advantage actor-critic algorithm is shown in Algorithm 2.

\begin{algorithm}[h]
  \caption{Training process}
  \begin{algorithmic}[1]
    \State
    {$Initialize\,\,paprameters\,\,of\,\,actor-critic\,\,network$};
    \While {$TRUE$}
    \State {$Update\,\,parameters\,\,of\,\,actor-critic\,\,network$};
    \State {$Learning\,\,new\,\, strategies\,\, from\,\, environment$};
    \State {$Update\,\,{s_t,a_t,r_t,s_{t+1}}$};
    \EndWhile
    \For {$i \in n$}
    \State {$Initialize\,\,agent\,\,a[i]$};
    \EndFor
    \While {$TRUE$}
    \For {$i \in n$}
    \State {$Training\,\,in\,\,environment$};
    \EndFor
    \State {$Update\,\,parameters\,\,of\,\,actor-critic\,\,network$};
    \For {$i \in n$}
    \State {$Update\,\,parameters\,\,of\,\,a[i]$};
    \EndFor
    \EndWhile
  \end{algorithmic}
\end{algorithm}

\section{Performance Evaluation and Analysis}\label{pa6}

\subsection{Parameter Setting}

Our simulation experiment is carried out in Anaconda3 + PyCharm3.6. We programmatically generate a physical network and 1,000 virtual networks, and save them in $.txt$ document. The physical network consists of 100 nodes and about 600 links. The CPU resource capacity and bandwidth resource capacity are randomly and evenly distributed between 50 units and 100 units. 1,000 virtual networks simulate Poisson process to reach the physical network, forming a continuous VNR process. Each virtual network randomly contains 2 to 12 virtual nodes, and each node has a 50\% probability to connect with each other. CPU resource demand and bandwidth resource demand of virtual nodes and virtual links are distributed randomly and evenly among 1 to 50 units. Specifically, the virtual network can reach 5 times in 100 time units, and the life cycle of each VNR is exponential distribution. The main parameters used in simulation experiment are shown in TABLE \ref{tab_2}.

\begin{table}
\centering
\caption{Parameter Setting}
\renewcommand\arraystretch{1.5}
\begin{tabular}{|p{35mm}|p{15mm}|}
\hline
Parameter & Value \\
\hline
Physical nodes & 100  \\
\hline
Physical links & 600 \\
\hline
CPU capacity & U[50,100] \\
\hline
Bandwidth capacity & U[50,100] \\
\hline
Virtual networks & 1,000 \\
\hline
Virtual nodes & U[2,12] \\
\hline
CPU requirements & U[1,50] \\
\hline
Bandwidth requirements & U[1,50] \\
\hline
\end{tabular}
\label{tab_2}
\end{table}

\subsection{Comparison Algorithms and Evaluation Indexes}

We select a heuristic VNE algorithm NodeRank \cite{2}, a RL based VNE algorithm MCST-VNE \cite{7} and a GCNN based VNE algorithm GCN-VNE \cite{13} as comparison algorithms. Then the performance of the algorithms is compared from two aspects: long-term average revenue of VNE and acceptance rate of VNR. The four algorithms, including our algorithm, are described and summarized in TABLE \ref{tab_3}.

\begin{table*}
\centering
\caption{Algorithm Idea Description}
\renewcommand\arraystretch{1.5}
\begin{tabular}{|p{20mm}|p{130mm}|}
\hline
Algorithm & Description \\
\hline
Our algorithm & The paper combines GCNN with RL algorithm, establishes a customized fitness matrix and fitness value, which aims to improve resource utilization, and improves the efficiency of the algorithm by model parallel training. \\
\hline
NodeRank \cite{2} & Sort physical nodes according to the importance of the nodes, and use the breadth-first search strategy for virtual link mapping. \\
\hline
MCST-VNE \cite{7} & The virtual node mapping is modeled as a Markov decision process, and MCST algorithm is used to specify the node mapping strategy. \\
\hline
GCN-VNE \cite{13} & A VNE algorithm based on DRL and GCNN, which trains the model through a designed multi-objective reward function. \\
\hline
\end{tabular}
\label{tab_3}
\end{table*}

In addition, we try to explore the impact of changing the resource capacity of physical network and the resource demand of VNRs on the flexibility of the algorithm. Specifically, we verify the revenue of the algorithm, the revenue cost ratio and the virtual network acceptance rate.

\subsection{Simulation Results and Performance Analysis}

\begin{figure}[!h]
\centering
\includegraphics[width=0.95\columnwidth]{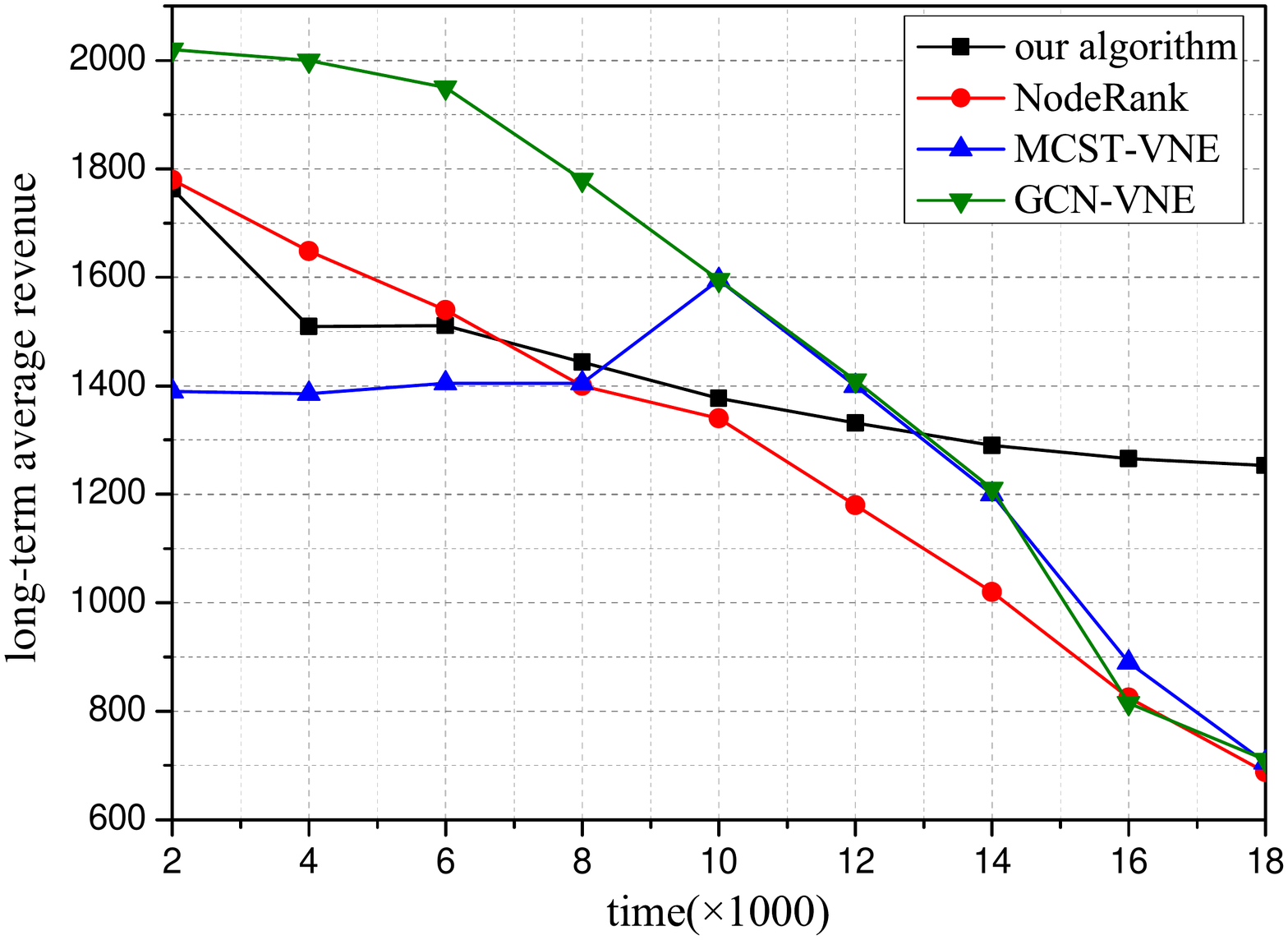}
\caption{Comparison with the long-term average revenue of other algorithms.}
\label{fig_4}
\end{figure}

We first compare the long-term average revenue of the four algorithms, and the result is shown in Fig. \ref{fig_4}. In the early stage of VNR, our algorithm performance is slightly worse than the other three algorithms. However, in the middle and late stages, our algorithm has a higher average revenue, indicating that our algorithm allocates physical network resources more reasonably, so that more resource consumption revenue can be obtained. From the experimental results, in the later stage of VNE, the long-term average revenue of our algorithm is 38.8\%, 22.5\% and 24\% higher than the other three algorithms, respectively. With the increase of time, the number of VNRs arriving at the physical network is increasing. The revenue calculation of VNE is shown in formula (9), which is determined by the consumption of physical resources. It should be noted that we calculate the long-term average revenue of VNE at time $t$ rather than the cumulative revenue. Therefore, when the number of VNRs is increasing, the number of physical network resources consumed is increasing, and the remaining available resources of underlying network are decreasing. At this time, the scale of VNRs it can carry becomes smaller, and the revenue of resource consumption will be correspondingly reduced. It can be seen from Fig. \ref{fig_4} that the trend of the curve is decreasing with time, so it is in line with the actual situation.

\begin{figure}[!h]
\centering
\includegraphics[width=0.95\columnwidth]{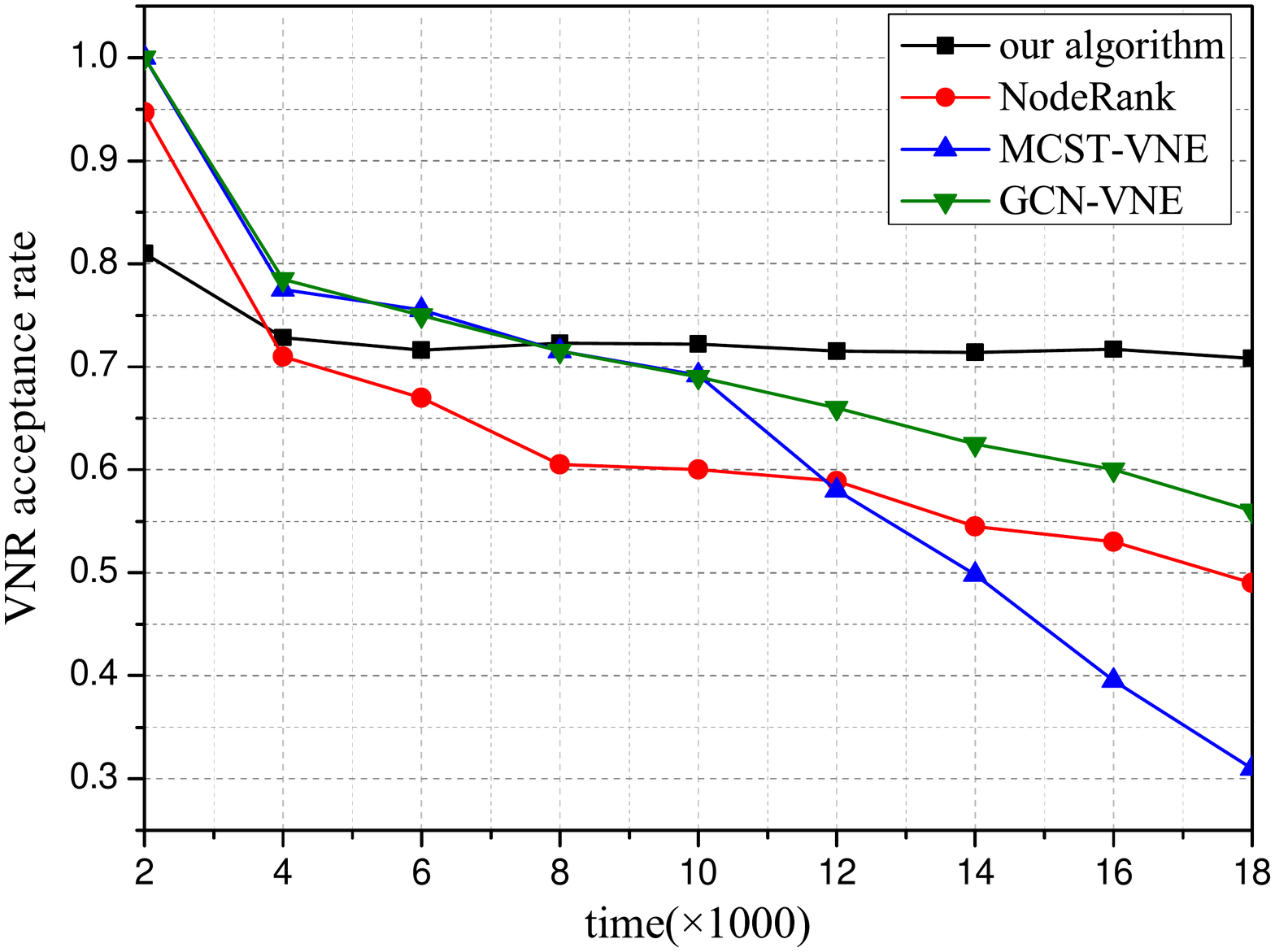}
\caption{Comparison with the VNR acceptance rate of other algorithms.}
\label{fig_5}
\end{figure}

We also compare the differences between the four algorithms in terms of virtual network acceptance rate, as shown in Fig. \ref{fig_5}. The change curve of VNR acceptance rate is similar to the revenue of VNE, but overall, the performance of our algorithm is better than the other three algorithms. After calculation, the acceptance rate of our algorithm is 15.2\%, 14.5\% and 2.7\% higher than the other three algorithms. An important reason is that we are concerned that VNE is a dynamic process. When selecting physical nodes for virtual nodes, we select those physical nodes with larger available resource capacity, so that the remaining physical resource capacity can also be used to carry other virtual nodes. On the one hand, it avoids resource fragmentation, and on the other hand, it improves the acceptance rate of VNRs. The acceptance rate of VNRs will continue to decrease over time. This is because physical network resources are continuously consumed, and its remaining resources can only carry a small portion of VNRs, so the acceptance rate of VNRs will decrease. Therefore, the changing trend of VNR acceptance rate is explainable and reasonable.

Generally speaking, our algorithm has better stability, and the performance of the other three algorithms has changed a lot. On the one hand, they give priority to physical nodes and physical links with large resource capacity for embedding, and ignore the reasonable degree of resource allocation. In addition, NodeRank is a VNE algorithm based on the ranking of node importance, which only considers the local importance of node attributes, and the flexibility of the algorithm will be limited by the manual physical node selection rules. It shows that the performance of VNE based on ML is better than VNE based on heuristics. MCST-VNE algorithm and GCN-VNE algorithm are based on ML algorithm, but they both ignore the feature that VNE is dynamic, which may cause waste in allocating physical resources for VNRs, which is the main reason that their performance is not optimal.

\begin{figure}[!h]
\centering
\includegraphics[width=0.95\columnwidth]{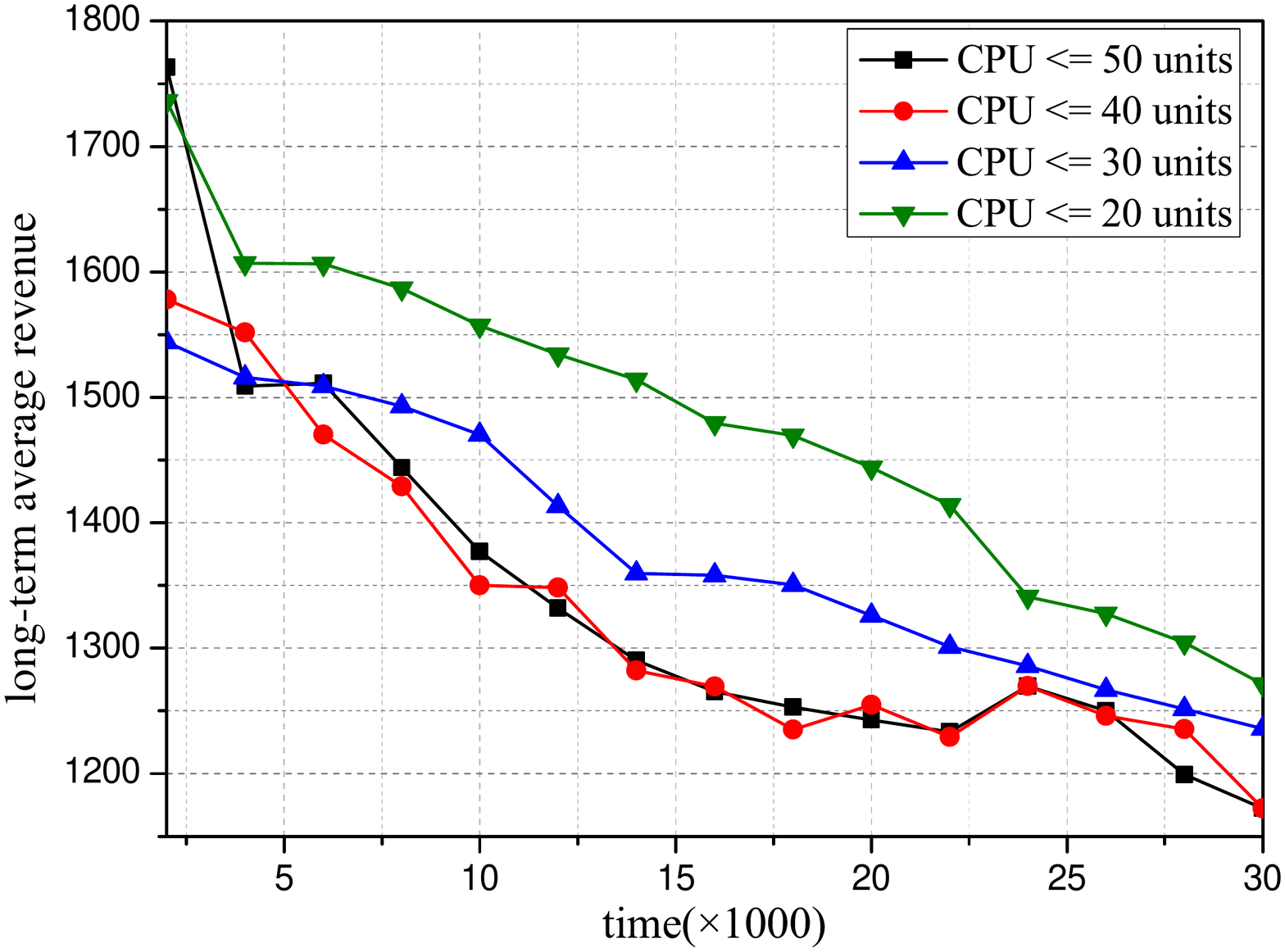}
\caption{Revenue test under different CPU resource requirements.}
\label{fig_6}
\end{figure}

\begin{figure}[!h]
\centering
\includegraphics[width=0.95\columnwidth]{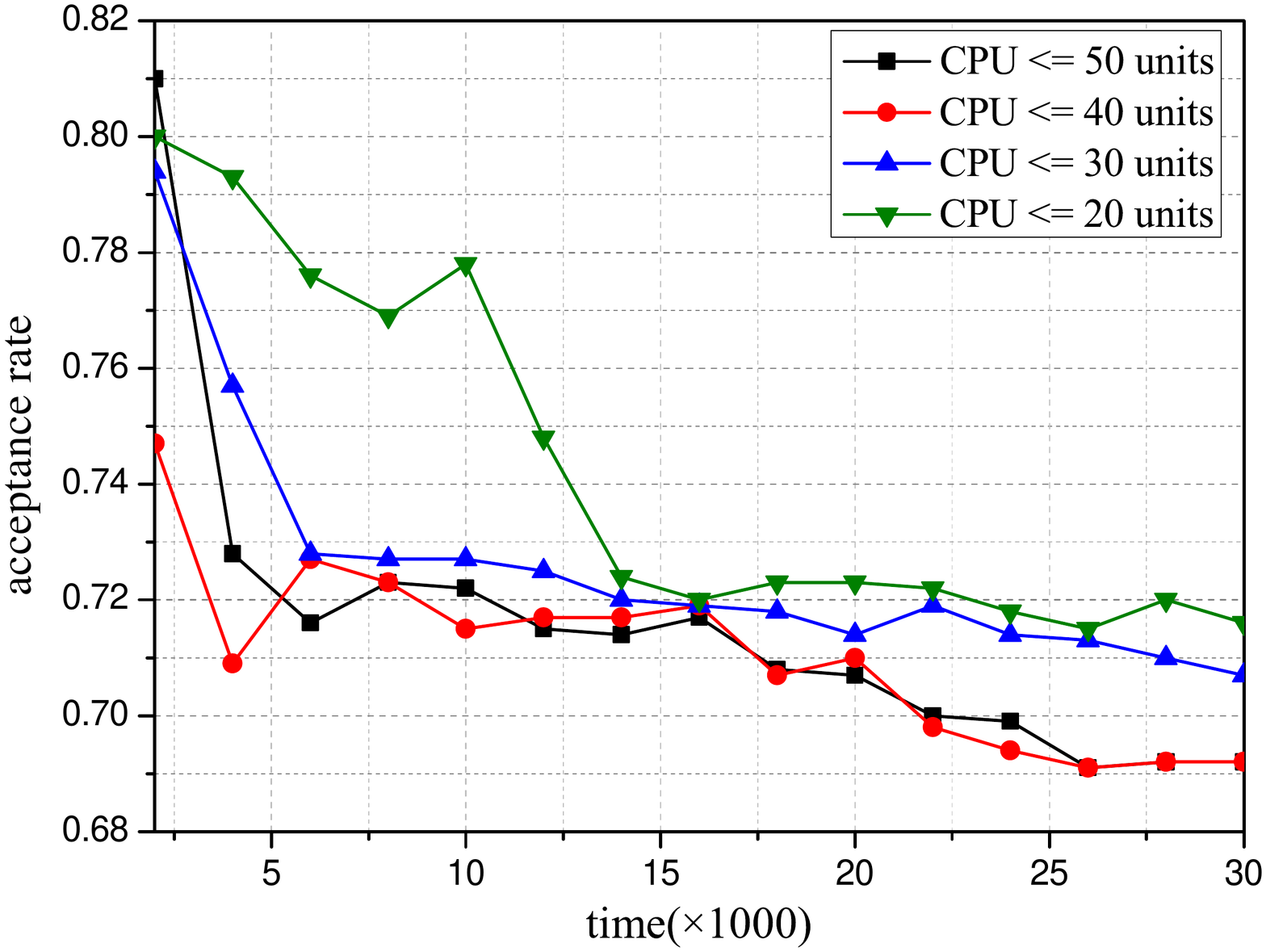}
\caption{Acceptance rate test under different CPU resource requirements.}
\label{fig_7}
\end{figure}

\begin{figure}[!h]
\centering
\includegraphics[width=0.95\columnwidth]{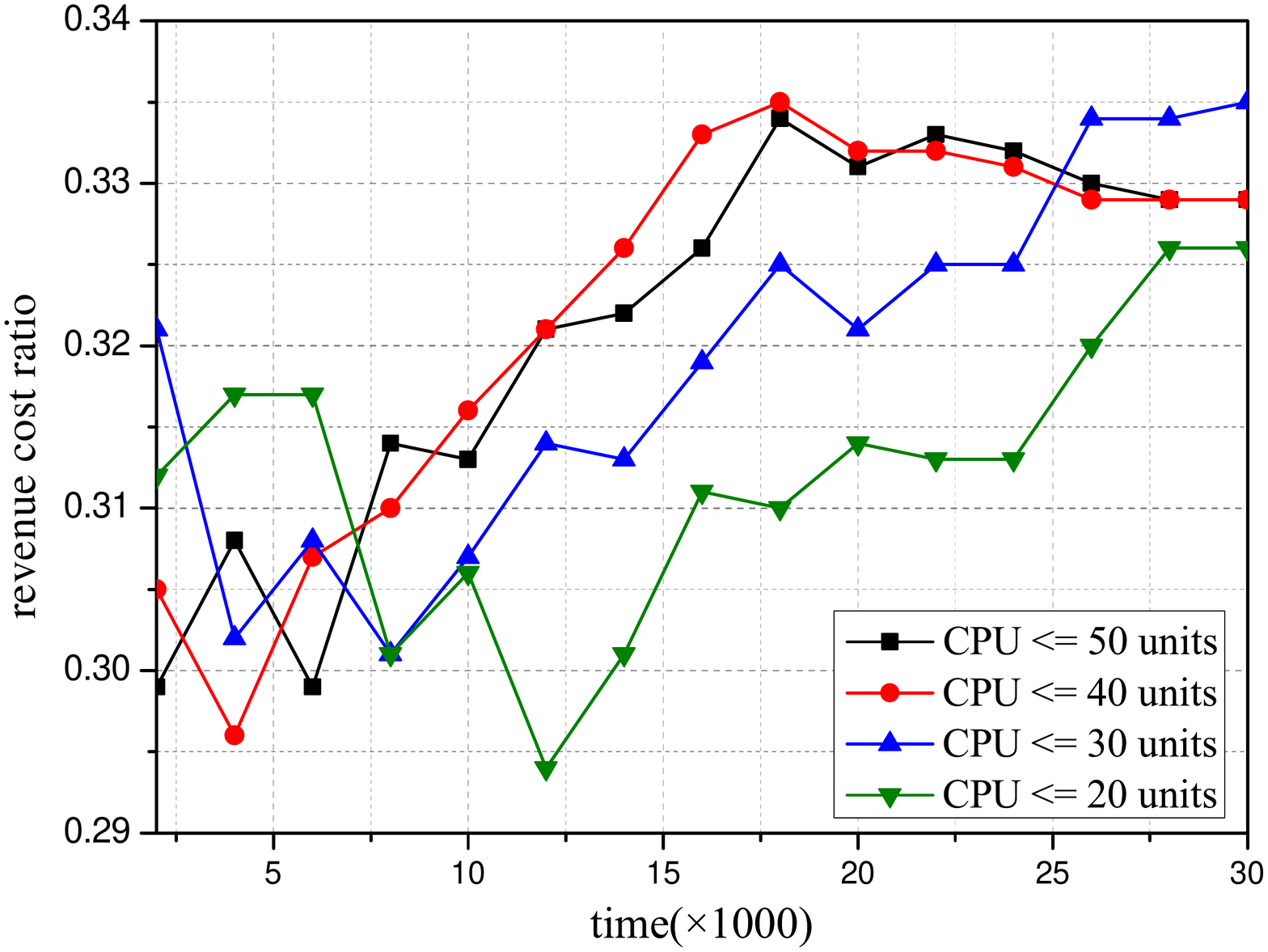}
\caption{Revenue cost ratio test under different CPU resource requirements.}
\label{fig_8}
\end{figure}

We try to verify the flexibility of the algorithm by changing the resource attributes of physical network or VNRs. Since the cost and revenue of VNE are closely related to nodes resources and links resources, in the parameter setting part, we set the same capacity range for CPU resources and bandwidth resources. So we use the method of changing the CPU resource requirement of virtual node and the fixed bandwidth resource requirement to explore the impact of the change of virtual network resource requirement attribute on the algorithm. The experimental results of algorithm flexibility verification are shown in Fig. \ref{fig_6} to Fig. \ref{fig_8}.

Fig. \ref{fig_6} shows the impact of changes in CPU resource requirements on long-term average revenue. Overall, the average revenue of the algorithm shows a downward trend under four different CPU resource requirements, which is caused by the decrease of available resources. A noticeable turning point is that when the CPU resource demand is less than or equal to 40 units, the long-term average revenue is the lowest. When the demand for CPU resources decreases, the long-term average revenue begins to rise again. It can be analyzed that when the CPU resource demand of the virtual node is less than or equal to 50 units, although the resource consumption of the virtual node is large, it can also bring higher revenue to the physical network. When the CPU resource requirement of virtual nodes is less than or equal to 30 or 20 units, although each VNR embedded in the physical network has low revenue, they can make up for the loss of revenue by increasing the number of VNE.


Fig. \ref{fig_7} and Fig. \ref{fig_8} show the results of the change of VNR acceptance rate and revenue cost ratio with the change of virtual network resource demand, respectively. A similar turning point occurs when the CPU resource requirement of virtual nodes is less than or equal to 40 units. The change curve of the revenue cost ratio is different from the other two. This is because the revenue cost ratio is not directly related to the size of the physical network resource capacity. It reflects the degree of resource utilization of the physical network and the profitability of the InPs. It can be inferred from the above experiments that changes in virtual link resource requirements can also produce similar effects. This shows that it is effective and realistic to explore the performance changes of the algorithm by changing the resource attributes of physical network or virtual network. Therefore, our algorithm has good flexibility.

\section{Conclusion}\label{pa7}

The VNE algorithm still faces several problems, including unreasonable constraints and goal setting, ignoring the time feasibility of the algorithm and not paying attention to the dynamic of the VNE process, which lead to the disadvantage of low flexibility of the existing VNE algorithm. This paper proposes a dynamic VNE algorithm based on RL and GCNN, taking into account the dynamics of VNE and the basic performance of the algorithm. We combine GCNN with RL algorithm, based on a self-defined fitness matrix and fitness value, to achieve an efficient dynamic VNE algorithm. This algorithm effectively reduces the degree of resource fragmentation by remapping VNRs. In RL, we use asynchronous advantage actor-critic algorithm for parallel training of the agent, which mainly includes key steps such as feature extraction, strategy generation and model training. In experimental phase, we first compare the proposed algorithm with other three representative algorithms. Afterwards, by changing the resource attributes of physical network and virtual network, it is more comprehensively demonstrate that the algorithm not only has satisfactory basic VNE performance, but also has good flexibility.

\ifCLASSOPTIONcaptionsoff
  \newpage
\fi


\begin{thebibliography}{1}

\bibitem{a1}
A. M. Alberti, G. D. Scarpioni, V. J. Magalh?es, A. Cerqueira S., J. J. P. C. Rodrigues and R. da Rosa Righi, ``Advancing NovaGenesis Architecture Towards Future Internet of Things,'' {\em IEEE Internet of Things Journal}, vol. 6, no. 1, pp. 215-229, Feb. 2019.

\bibitem{z1}
P. Zhang, C. Wang, G. S. Aujla, N. Kumar and M. Guizani, ``IoV Scenario: Implementation of a Bandwidth Aware Algorithm in Wireless Network Communication Mode,'' {\em IEEE Transactions on Vehicular Technology}, vol. 69, no. 12, pp. 15774-15785, Dec. 2020.

\bibitem{a2}
S. Cherrared, S. Imadali, E. Fabre, G. G?ssler and I. G. B. Yahia, ``A Survey of Fault Management in Network Virtualization Environments: Challenges and Solutions,'' {\em IEEE Transactions on Network and Service Management}, vol. 16, no. 4, pp. 1537-1551, Dec. 2019.

\bibitem{a3}
S. Nazemi, K. K. Leung and A. Swami, ``Distributed Optimization Framework for In-Network Data Processing,'' {\em IEEE/ACM Transactions on Networking}, vol. 27, no. 6, pp. 2432-2443, Dec. 2019.

\bibitem{z2}
P. Zhang, X. Pang, Y. Bi, H. Yao, H. Pan and N. Kumar, ``DSCD: Delay Sensitive Cross-Domain Virtual Network Embedding Algorithm,'' {\em IEEE Transactions on Network Science and Engineering}, vol. 7, no. 4, pp. 2913-2925, Oct.-Dec. 2020.

\bibitem{a4}
H. Cao, H. Zhu and L. Yang, ``Collaborative attributes and resources for single-stage virtual network mapping in network virtualization,'' {\em Journal of Communications and Networks}, vol. 22, no. 1, pp. 61-71, Feb. 2020.

\bibitem{jcx1}
C. Jiang, Y. Chen, K. J. R. Liu and Y. Ren, ``Renewal-Theoretical Dynamic Spectrum Access in Cognitive Radio Network with Unknown Primary Behavior,'' {\em IEEE Journal on Selected Areas in Communications}, vol. 31, no. 3, pp. 406-416, 2013.

\bibitem{jcx2}
C. Jiang, Y. Chen, Y. Gao and K. J. R. Liu, ``Joint Spectrum Sensing and Access Evolutionary Game in Cognitive Radio Networks,'' {\em IEEE Transactions on Wireless Communications}, vol. 12, no. 5, pp. 2470-2483, 2013.

\bibitem{jcx3}
X. Zhu, C. Jiang, L. Kuang, N. Ge and J. Lu, ``Non-orthogonal Multiple Access Based Integrated Terrestrial-Satellite Networks,'' {\em IEEE Journal on Selected Areas in Communications}, vol. 35, no. 10, pp. 2253-2267, Oct. 2017.

\bibitem{a5}
R. Chai, D. Xie, L. Luo and Q. Chen, ``Multi-Objective Optimization-Based Virtual Network Embedding Algorithm for Software-Defined Networking,'' {\em IEEE Transactions on Network and Service Management}, vol. 17, no. 1, pp. 532-546, Mar. 2020.

\bibitem{a6}
M. Lu, Y. Gu and D. Xie, ``A Dynamic and Collaborative Multi-Layer Virtual Network Embedding Algorithm in SDN Based on Reinforcement Learning,'' {\em IEEE Transactions on Network and Service Management}, vol. 17, no. 4, pp. 2305-2317, Dec. 2020.

\bibitem{z3}
P. Zhang, C. Wang, C. Jiang and A. Benslimane, ``Security-Aware Virtual Network Embedding Algorithm based on Reinforcement Learning,'' {\em IEEE Transactions on Network Science and Engineering, early access}, pp. 1-1, 2020, doi: 10.1109/TNSE.2020.2995863.

\bibitem{a7}
H. Afifi and H. Karl, ``Reinforcement Learning for Virtual Network Embedding in Wireless Sensor Networks,'' {\em 2020 16th International Conference on Wireless and Mobile Computing, Networking and Communications (WiMob), Thessaloniki, Greece}, pp. 123-128, 2020.

\bibitem{a8}
I. Budhiraja, N. Kumar and S. Tyagi, ``Deep-Reinforcement-Learning-Based Proportional Fair Scheduling Control Scheme for Underlay D2D Communication,'' {\em IEEE Internet of Things Journal}, vol. 8, no. 5, pp. 3143-3156, Mar. 2021.

\bibitem{a9}
Y. Li, R. Chen, Y. Zhang and H. Li, ``A CNN-GCN Framework for Multi-Label Aerial Image Scene Classification,'' {\em IGARSS 2020 - 2020 IEEE International Geoscience and Remote Sensing Symposium, Waikoloa, HI, USA}, pp. 1353-1356, 2020.

\bibitem{z4}
P. Zhang, C. Wang, C. Jiang and Z. Han, ``Deep Reinforcement Learning Assisted Federated Learning Algorithm for Data Management of IIoT,'' {\em IEEE Transactions on Industrial Informatics, early access}, pp. 1-1, 2021, doi: 10.1109/TII.2021.3064351.

\bibitem{1}
M. Chowdhury, M. R. Rahman and R. Boutaba, ``ViNEYard: Virtual Network Embedding Algorithms With Coordinated Node and Link Mapping,'' {\em IEEE/ACM Transactions on Networking}, vol. 20, no. 1, pp. 206-219, Feb. 2012.

\bibitem{2}
X. Cheng, S. Su, Z. Zhang, H. Wang, F. Yang, Y. Luo and J. Wang, ``Virtual network embedding through topology-aware node ranking,'' {\em Computer Communication Review}, vol. 41, no. 2, pp. 38-47, Apr. 2011.

\bibitem{3}
P. Zhang, H. Yao and Y. Liu, ``Virtual Network Embedding Based on Computing, Network, and Storage Resource Constraints,'' {\em IEEE Internet of Things Journal}, vol. 5, no. 5, pp. 3298-3304, Oct. 2018.

\bibitem{m2}
J. Rubio-Loyola, C. Aguilar-Fuster, G. Toscano-Pulido, R. Mijumbi and J. Serrat-Fernandez, ``Enhancing Metaheuristic-Based Online Embedding in Network Virtualization Environments,'' {\em IEEE Transactions on Network and Service Management}, vol. 15, no. 1, pp. 200-216, Mar. 2018.

\bibitem{4}
A. Song, W. -N. Chen, T. Gu, H. Yuan, S. Kwong and J. Zhang, ``Distributed Virtual Network Embedding System With Historical Archives and Set-Based Particle Swarm Optimization,'' {\em IEEE Transactions on Systems, Man, and Cybernetics: Systems}, vol. 51, no. 2, pp. 927-942, Feb. 2021.

\bibitem{5}
P. Zhang, Y. Hong, X. Pang and C. Jiang, ``VNE-HPSO: Virtual Network Embedding Algorithm Based on Hybrid Particle Swarm Optimization,'' {\em IEEE Access}, vol. 8, pp. 213389-213400, 2020.

\bibitem{m1}
A. Song, W. -N. Chen, T. Gu, H. Zhang and J. Zhang, ``A Constructive Particle Swarm Optimizer for Virtual Network Embedding,'' {\em IEEE Transactions on Network Science and Engineering}, vol. 7, no. 3, pp. 1406-1420, Jul.-Sep. 2020.

\bibitem{6}
H. Yao, S. Ma, J. Wang, P. Zhang, C. Jiang and S. Guo, ``A Continuous-Decision Virtual Network Embedding Scheme Relying on Reinforcement Learning,'' {\em IEEE Transactions on Network and Service Management}, vol. 17, no. 2, pp. 864-875, Jun. 2020.

\bibitem{7}
S. Haeri and L. Trajkovic, ``Virtual Network Embedding via Monte Carlo Tree Search,'' {\em IEEE Transactions on Cybernetics}, vol. 48, no. 2, pp. 510-521, Feb. 2018.

\bibitem{8}
Y. Yuan, Z. Tian, C. Wang, F. Zheng and Y. Lv, ``A Q-learning-based approach for virtual network embedding in data center,'' {\em Neural Computing and Applications}, vol. 32, no. 7, pp. 1995-2004, Apr. 2020.

\bibitem{9}
H. Yao, X. Chen, M. Li, P. Zhang and L. Wang, ``A novel reinforcement learning algorithm for virtual network embedding," {\em Neurocomputing}, vol. 284, pp. 1-9, Apr. 2018.

\bibitem{10}
M. Dolati, S. B. Hassanpour, M. Ghaderi and A. Khonsari, ``DeepViNE: Virtual Network Embedding with Deep Reinforcement Learning,'' {\em IEEE INFOCOM 2019 - IEEE Conference on Computer Communications Workshops (INFOCOM WKSHPS), Paris, France}, pp. 879-885, 2019.

\bibitem{11}
C. Wang, R. S. Batth, P. Zhang, G. S. Aujla, Y. Duan and L. Ren, ``VNE solution for network differentiated QoS and security requirements: from the perspective of deep reinforcement learning,'' {\em Computing}, vol. 103, no. 6, pp. 1061-1083, 2021.

\bibitem{12}
F. Habibi, M. Dolati, A. Khonsari and M. Ghaderi, ``Accelerating Virtual Network Embedding with Graph Neural Networks,'' {\em 2020 16th International Conference on Network and Service Management (CNSM), Izmir, Turkey}, pp. 1-9, 2020.

\bibitem{13}
Z. Yan, J. Ge, Y. Wu, L. Li and T. Li, ``Automatic Virtual Network Embedding: A Deep Reinforcement Learning Approach With Graph Convolutional Networks," {\em IEEE Journal on Selected Areas in Communications}, vol. 38, no. 6, pp. 1040-1057, Jun. 2020.

\bibitem{14}
C. Zhao and B. Parhami, ``Virtual Network Embedding Through Graph Eigenspace Alignment,'' {\em IEEE Transactions on Network and Service Management}, vol. 16, no. 2, pp. 632-646, Jun. 2019.

\bibitem{15}
T. Ghazar and N. Samaan, ``A Hierarchical Approach for Efficient Virtual Network Embedding Based on Exact Subgraph Matching,'' {\em 2011 IEEE Global Telecommunications Conference - GLOBECOM 2011, Houston, TX, USA}, pp. 1-6, 2011.

\bibitem{16}
C. K. Dehury and P. K. Sahoo, ``DYVINE: Fitness-Based Dynamic Virtual Network Embedding in Cloud Computing,'' {\em IEEE Journal on Selected Areas in Communications}, vol. 37, no. 5, pp. 1029-1045, May 2019.

\bibitem{17}
M. Lu, Y. Gu and D. Xie, ``A Dynamic and Collaborative Multi-Layer Virtual Network Embedding Algorithm in SDN Based on Reinforcement Learning,'' {\em IEEE Transactions on Network and Service Management}, vol. 17, no. 4, pp. 2305-2317, Dec. 2020.

\bibitem{18}
H. Cao, S. Wu, G. S. Aujla, Q. Wang, L. Yang and H. Zhu, ``Dynamic Embedding and Quality of Service-Driven Adjustment for Cloud Networks,'' {\em IEEE Transactions on Industrial Informatics}, vol. 16, no. 2, pp. 1406-1416, Feb. 2020.

\bibitem{19}
M. Defferrard, X. Bresson, and P. Vandergheynst, ``Convolutional neural networks on graphs with fast localized spectral filtering,'' {\em Advances in Neural Information Processing Systems}, vol. 0, pp. 3844-3852, 2016.

\end{thebibliography}
\end{document}